\documentclass[a4paper]{article}

\usepackage{amsmath}

\DeclareMathOperator*{\sgn}{\mathrm{sgn}}

\usepackage{INTERSPEECH2022}
\usepackage{algorithm}
\usepackage{algpseudocode}
\usepackage{setspace}
\usepackage{multirow}
\usepackage[caption=false]{subfig}

\title{SMP-PHAT: Lightweight DoA Estimation by Merging Microphone Pairs}
\name{Fran\c{c}ois Grondin, Marc-Antoine Maheux, Jean-Samuel Lauzon, \\Jonathan Vincent, Fran\c{c}ois Michaud}
\address{
  Universit\'e de Sherbrooke, Sherbrooke (Qu\'ebec), Canada}
\email{\{francois.grondin2, marc-antoine.maheux, jean-samuel.lauzon\\jonathan.vincent2, francois.michaud\}@usherbrooke.com}

\begin{document}

\maketitle
\begin{abstract}
This paper introduces SMP-PHAT, which performs direction of arrival (DoA) of sound estimation with a microphone array by merging pairs of microphones that are parallel in space.
This approach reduces the number of pairwise cross-correlation computations, and brings down the number of flops and memory lookups when searching for DoA. Experiments on low-cost hardware with commonly used microphone arrays show that the proposed method provides the same accuracy as the former SRP-PHAT approach, while reducing the computational load by 39\% in some cases.
\end{abstract}

\section{Introduction}

Most speech driven intelligent devices (e.g. smart speakers, tablets, cell phones) use multiple microphones to enhance distant speech prior to speech recognition tasks \cite{barker2015third,pertila2018multichannel}.
While some advanced recognition tasks can be performed on the cloud, multi-channel enhancing steps are usually done locally on the devices.
Speech enhancement methods can benefit from sound source localization (SSL), which aims at estimating the direction of arrival (DoA) of the target sound source.
DoA can help to solve the permutation ambiguity by providing additional parameters for deep clustering \cite{wang2018multi,drude2019unsupervised}, and guide time-frequency mask generation for beamforming \cite{grondin2020gev,wood2017blind,pertila2015distant}.
DoA estimation can rely on beamforming approaches, subspace decomposition methods or deep learning-based techniques.

Steered-Response Power beamforming with Phase Transform (SRP-PHAT) consists in pointing a beamformer at each potential direction of arrival (DoA) of sound and compute the corresponding power \cite{do2007real}.
The direction with the most power is considered as the most likely DoA for the sound source.
SRP-PHAT is often performed using Time Difference of Arrival (TDoA) between each pair of microphones using the Generalized Cross-Correlation with Phase Transform (GCC-PHAT) method \cite{valin2004localization,valin2007robust}.
GCC-PHAT is an appealing method, as it is robust to reverberation, and exploits the efficient Inverse Fast Fourier Transform (IFFT) algorithm \cite{brandstein1997robust}.
The Phase Transform however remains sensitive to additive noise, and time-frequency masking to using signal processing \cite{cohen2001speech, grondin2015time, grondin2016noise} or deep-learning \cite{pertila2017robust, wang2018robust} based methods can improve DoA estimation robustness for a given class of sound signals (e.g. speech).
SSL with beamforming is well-suited for low-cost hardware, but the complexity increases significantly as the number of microphones increases and the number of potential DoAs gets large \cite{grondin2013manyears, grondin2019lightweight, grondin2021odas}. 

Estimation of signal parameters via rotational invariance technique (ESPRIT) \cite{han2005esprit} and Multiple Signal Classification (MUSIC) \cite{schmidt1986multiple} exploit the signal and noise subspaces, respectively.
MUSIC was formerly used with narrow-band signals, and then adapted to broadband signals such as speech, and rely on Standard Eigenvalue Decomposition (SEVD-MUSIC) \cite{ishi2009evaluation}.
Generalized Eigenvalue Decomposition with MUSIC (GEVD-MUSIC) is also proposed to deal with target speech in negative signal-to-noise ratio (SNR) scenarios \cite{nakamura2009intelligent,nakamura2011intelligent,nakadai2012robot}.
It was also shown that using Generalized Singular Value Decomposition with MUSIC (GSVD-MUSIC) improves accuracy as it ensures orthogonality of the noise subspace bases \cite{nakamura2012real}.
These subspace decomposition methods however rely on eigenvalue or singular value decompositions, which involve a significant amount of computations, and make real-time processing on low-cost hardware challenging.

Machine learning can also be used to estimate the DoA for a given microphone array geometry.
For instance, Convolutional Neural Network (CNN) can be applied to phase maps trained using noise signals to estimate the DoA of a sound source for a linear microphone array \cite{chakrabarty2017broadband,chakrabarty2019multi}.
It is also possible to exploit Convolutional Recurrent Neural Networks (CRNN) to estimate simultaneously the DoA and associated class of signal for a 4-microphone array \cite{adavanne2018sound,adavanne2018direction}.
Another strategy aims at breaking down the problem to estimate TDoAs for each pair of microphones using a CRNN, and then combining the results to estimate the DoA to adapt to different microphone array geometries \cite{grondin2019sound}.
A neural network can also be trained on a generic geometry and adapted to new array structures using small amount of data \cite{li2018online}.
While machine learning-based SSL provides accurate results, it also requires multi-channel training data and involves a significant amount of parameters, which can be challenging when performing inference on low-cost hardware.

As SRP-PHAT remains appealing with low-cost hardware, this paper proposes a improved version, called SMP-PHAT, for Steered response power by Merging Pairs with PHAse Transform.
This method aims to reduce the number of computations involved in SRP-PHAT when dealing with numerous microphones and scanning space in 3-D.
The proposed approach is particularly well-suited for commercial microphone arrays with small aperture (a few centimeters) and symmetrical geometries.
Section \ref{sec:srpphat} first describes the SRP-PHAT original algorithm, and then section \ref{sec:smpphat} introduces the proposed SMP-PHAT method.
Section \ref{sec:results} present the results, which demonstrate that SMP-PHAT can reduce the number of computations by 40\% in some cases, and speed benchmarks are provided for the Raspberry Pi 4 \cite{mcmanus2021raspberry} and Zero \cite{grimmett2016getting} low-cost devices.
Section \ref{sec:conclusion} concludes with final remarks and future work.

\section{SRP-PHAT}
\label{sec:srpphat}

An array consists of $M \in \mathbb{N}$ microphones, where each microphone $m \in \{1, 2, \dots, M\}$ is at position $\mathbf{x}_m \in \mathbb{R}^3$.
Each pair is indexed by $p \in \{1, 2, \dots, P\}$, where $P=M(M-1)/2$ is the number of pairs.
The Short-Time Fourier Transform with a frame size of $N \in \mathbb{N}$ samples is usually performed on each microphone $m$ to obtain the time-frequency representation, denoted as $X_m[t,f] \in \mathbb{C}$, where $t \in \mathbb{N}$ stands for the frame index, and $f \in \{0, 1, \dots, N/2\}$ stands for the frequency bin indexes.
The cross-correlation in the frequency domain between microphones $u$ and $v \in \{1, 2, \dots, M\}$ that correspond to pair $p$ is usually computed over a window of frames as follows:
\begin{equation}
    C_p[f] = \sum_{t}{X_u[t,f]X_v[t,f]^*},
\end{equation}
where $(\dots)^*$ stands for the complex conjugate.
The cross-correlation can be normalized using the Phase Transform as follows:
\begin{equation}
    R_p[f] = \frac{C_p[f]}{|C_p[f]|}.
\end{equation}

For each pair of microphone, the cross-correlation in the time domain is computed using the Generalized Cross-Correlation with Phase Transform (GCC-PHAT):
\begin{equation}
r_p[\tau] = \sum_{f=0}^{N/2}{R_p[f] e^{j 2\pi f\tau/N}}
\label{eq:r}
\end{equation}
where the expression $j$ stands for the imaginary number $\sqrt{-1}$.
When the TDoA $\tau$ is an integer, the expression in (\ref{eq:r}) can be computed using an Inverse Fast Fourier Transform (IFFT), which speeds up computation significantly.
To cope with TDoA rounding introduced by the IFFT, the frame size can be increased with interpolation by a factor of $k \in \{1, 2, 4, \dots\}$.
Given the exact microphone array geometry is known, the expression $\mathbf{d}_p = \mathbf{x}_u - \mathbf{x}_v$ stands for the difference between the positions $\mathbf{x}_u \in \mathbb{R}^3$ and $\mathbf{x}_v \in \mathbb{R}^3$ of the microphones $(u,v)$ in pair $p$.
For each potential direction of arrival (DoA) and pair of microphones, a time difference of arrival (TDoA) can be estimated offline: 
\begin{equation}
\tau_p[i] = \frac{1}{k}\left\lfloor k\left(\frac{f_S}{c}\right)\mathbf{d}_p \cdot \mathbf{u}_i \right\rceil
\end{equation}
where $f_S \in \mathbb{N}$ stands for the sample rate in samples/sec and $c \in \mathbb{R}^+$ for the speed of sound in m/sec.
The unit vector $\mathbf{u}_i$ stands for the DoA, where $i \in \{1, 2, \dots, I\}$ for a total of $I$ DoAs.
The symbol $(\cdot)$ stands for the dot product and the operator $\lfloor\dots\rceil$ rounds to the closest integer.

Algorithm \ref{alg:srpphat} shows the steps performed to find the most likely DoA.
The cross-correlation is computed for each pair of microphones, and the power is computed for each DoA, for a total of $I$ directions.
The predicted DoA is stored in the expression $\mathbf{u}^*$.
This approach involves $P$ IFFTs, $PI$ lookups and $PI$ additions.
This can become challenging for real-time processing, especially for a large number of potential DoAs and when the number of microphones increases (the complexity corresponds to $\mathcal{O}(M^2)$, as $P = M(M-1)/2$).

\begin{algorithm}
\caption{Online SRP-PHAT}\label{alg:srpphat}
\begin{algorithmic}
\For{$p \in \{1, 2, \dots, P\}$}
    \State $r_p[\tau] \gets \mathrm{IFFT}\{R_p[f]\}$
\EndFor
\State $E_{max} \gets 0$
\State $i_{max} \gets 0$
\For{$i \in \{1, 2, \dots, I\}$}
    \State $E \gets 0$
    \For{$p \in \{1, 2, \dots, P\}$}
        \State $E \gets E + r_p[k\tau_p[i]]$
    \EndFor
    \If{$E > E_{max}$}
        \State $i_{max} \gets i$
        \State $E_{max} \gets E$
    \EndIf
\EndFor
\State $\mathbf{u}^* \gets \mathbf{u}_{i_{max}}$
\end{algorithmic}
\end{algorithm}

\begin{figure}[!ht]
    \centering
    \includegraphics[width=\linewidth]{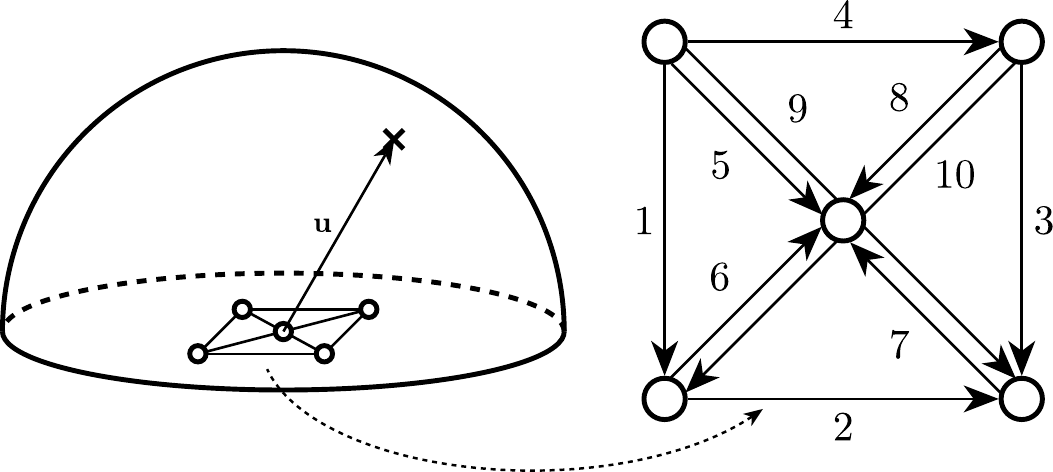}
    \caption{Example of a microphone array geometry with 5 microphones and 10 pairs, and potential DoAs on a half-sphere made of 1321 discrete directions}
    \label{fig:example}
\end{figure}

For instance, Figure \ref{fig:example} shows a microphone array with four microphones at each corner of a square shape and fifth one in the center.
In this configuration, there are 10 pairs of microphones, and the DoAs correspond to the points on a half-sphere over the array with unit norm (generated from a icosahedron as in [?] with a resolution of approximately $1^o$), for a total of $I=1321$ directions.
In this example, $10$ cross-correlations are computed using $10$ IFFTs, and then each DoA is associated to $10$ lookups and addition, for a total of $13210$ lookups and $13210$ additions.

\section{SMP-PHAT}
\label{sec:smpphat}

The proposed method, named SMP-PHAT for Steered response power by Merging Pairs with PHAse Transform, aims to reduce the number of computations when there is a large number of microphones and there is some symmetry in the shape of the array. 
In the former example in Figure \ref{fig:example}, it is possible to define by inspection the sets $\mathcal{B}_1 = \{1,3\}$, $\mathcal{B}_2 = \{2,4\}$, $\mathcal{B}_3 = \{5,7\}$, $\mathcal{B}_4 = \{6,8\}$, $\mathcal{B}_5 = \{9\}$ and $\mathcal{B}_6 = \{10\}$, where each set holds the index of one or many pairs that are equidistant and parallel.
Algorithm \ref{alg:merging} proposes a more formal way to perform merging of pairs with identical orientation and distance.
The algorithm starts with the first pair, and combines the other pairs that can be merged.
For two pairs to be merged, the magnitude of their dot product and the norm of one of the pair must be equal (or the absolute value of their difference should be below a small threshold $\epsilon = 10^{-4}$ to account for floating point errors).
These pairs are then removed from the initial set, and this is repeated until all pairs are assigned to a new set.
At the end, there are $Q \leq P$ sets, each one denoted as $\mathcal{B}_q$, where $q \in \{1, 2, \dots, Q\}$ stands for the index.
This task can easily be performed offline and the results stored in memory.
Note that this approach relies on the farfield assumption, which is suitable here for microphone arrays with small aperture (a few centimeters) and a sound source a few meters away.

\begin{algorithm}
\caption{Offline pair merging plan for SMP-PHAT}\label{alg:merging}
\begin{algorithmic}
\State $Q \gets 0$
\State $\mathcal{A} \gets \{1, 2, \dots, P\}$
\For{$p \in \mathcal{A}$}
    \State $Q \gets Q + 1$
    \State $\mathcal{B}_Q \gets \{p\}$
    \State $\mathcal{A} \gets \mathcal{A} \setminus \{p\}$
    \For{$p' \in \mathcal{A}$}
        \If{$\left||\mathbf{d}_p \cdot \mathbf{d}_{p'}|-\lVert\mathbf{d}_{p}\rVert_2\lVert\mathbf{d}_{p'}\rVert_2\right| < \epsilon$}
            \If{$\left|\lVert\mathbf{d}_{p}\rVert_2-\lVert\mathbf{d}_{p'}\rVert_2\right| < \epsilon$}
                \State $\mathcal{B}_Q \gets \mathcal{B}_Q \cup \{p'\}$
                \State $\mathcal{A} \gets \mathcal{A} \setminus \{p'\}$
            \EndIf
        \EndIf
    \EndFor
\EndFor
\end{algorithmic}
\end{algorithm}

To make notation simpler, we define a reference pair in each set $\mathcal{B}_q$ that corresponds to the pair with the lowest index:
\begin{equation}
    \bar{q} = \min\{\mathcal{B}_q\}.
\end{equation}

When two pairs are in the same set ($p,p' \in \mathcal{B}_q$), the magnitude of their TDoAs for every potential DoA $\mathbf{u}_i$ are identical, such that $|\tau_{p'}[i]| = |\tau_{p}[i]|$.
When two pairs are parallel but in opposite direction, the magnitude is the same but the sign is different.
If we denote $\hat{r}$ as the time reversed signal $r$ ($\hat{r}_p[\tau] = r_p[-\tau]$), we can define a new signal that sums all the cross-correlation in each set:
\begin{equation}
    s_q[\tau] = \sum_{p \in \mathcal{B}_q}{
    \begin{cases}
        r_{\bar{q}}[\tau] & \mathrm{if}\ \ \sgn\{\mathbf{d}_p \cdot \mathbf{d}_{\bar{q}}\} = +1\\
        \hat{r}_{\bar{q}}[\tau] & \mathrm{if}\ \ \sgn\{\mathbf{d}_p \cdot \mathbf{d}_{\bar{q}}\} = -1 \\
    \end{cases}},
    \label{eq:sq}
\end{equation}
where $\sgn\{\dots\}$ is the signum function.
Time reversal can be expressed as the complex conjugate of the signal in the frequency domain, such that:
\begin{equation}
    \hat{R}_p[f] = (R_p[f])^*,
\end{equation}
and therefore (\ref{eq:sq}) becomes:
\begin{equation}
    S_q[f] = \sum_{p \in \mathcal{B}_q}{
    \begin{cases}
        R_p[f] & \mathrm{if}\ \ \sgn\{\mathbf{d}_p \cdot \mathbf{d}_{\bar{q}}\} = +1\\
        \hat{R}_p[f] & \mathrm{if}\ \ \sgn\{\mathbf{d}_p \cdot \mathbf{d}_{\bar{q}}\} = -1\\
    \end{cases}
    },
    \label{eq:Sq}
\end{equation}
which involves adding $N/2+1$ complex numbers, and corresponds to $N+2$ real additions.
It is therefore possible to compute $S_q[f]$ in the frequency domain, and perform a unique IFFT to obtain $s_q[\tau]$:
\begin{equation}
    s_q[\tau] = \sum_{f=0}^{N/2}{S_q[f] e^{j 2\pi f\tau/N}}.
\end{equation}

Algorithm \ref{alg:proposed} exploits these useful properties.
In this approach, the number of IFFTs reduces to $Q$, the number of lookups to $QI$ and the number of additions to $QI+(N+2)(P-Q)$.
Note that the extra $(N+2)(P-Q)$ additions come from computing the expression $S_q[f]$ for each $q$.
The reduction in the number of computation thus equals $(P-Q)/P$ for the IFFT and lookups, and corresponds to $PI/(QI+(N+2)(P-Q))$ for the additions.
Table \ref{tab:computations} summarizes the number of computations with the SRP-PHAT and SMP-PHAT.

\begin{algorithm}
\caption{Online SMP-PHAT}\label{alg:proposed}
\begin{algorithmic}
\For{$q \in \{1, 2, \dots, Q\}$}
    \State Compute $S_q[f]\ \forall\ f$
    \State $s_q[\tau] \gets \mathrm{IFFT}\{S_q[f]\}$
\EndFor
\State $E_{max} \gets 0$
\State $i_{max} \gets 0$
\For{$i \in \{1, 2, \dots, I\}$}
    \State $E \gets 0$
    \For{$q \in \{1, 2, \dots, Q\}$}
        \State $E \gets E + s_q[\tau_{\bar{q}}[i]]$
    \EndFor
    \If{$E > E_{max}$}
        \State $i_{max} \gets i$
        \State $E_{max} \gets E$
    \EndIf
\EndFor
\State $\mathbf{u}^* \gets \mathbf{u}_{i_{max}}$
\end{algorithmic}
\end{algorithm}

\begin{table}[!ht]
    \centering
    \caption{Number of computations (SRP-PHAT and SMP-PHAT)}
    \renewcommand{\arraystretch}{1.4}
    \begin{tabular}{c|ccc}
    \hline
    \hline
        Method & IFFTs & Lookups & Additions \\
    \hline
        SRP & $PI$ & $PI$ & $PI$ \\
        SMP & $QI$ & $QI$ & $QI + (N+2)(P-Q)$ \\
        $\Delta$ & $\frac{P-Q}{Q}$ & $\frac{P-Q}{Q}$ & $\frac{PI-QI+(N+2)(P-Q)}{QI+(N+2)(P-Q)}$ \\
    \hline
    \hline
    \end{tabular}
    \vspace{-5pt}
    \label{tab:computations}
\end{table}

For example, using the configuration in Figure \ref{fig:example}, we set $Q=6$, such that the number of IFFTs drops to $6$ (reduction by 40\%), and the number of lookups to $7926$ (reduction by 40\%).
Assuming a frame size of $N=512$, the number of additions goes down to $9982$ (reduction by 24\%).

\section{Results}
\label{sec:results}

\begin{figure}
    \centering
    \subfloat[ReSpeaker USB]{\includegraphics[width=0.4\linewidth]{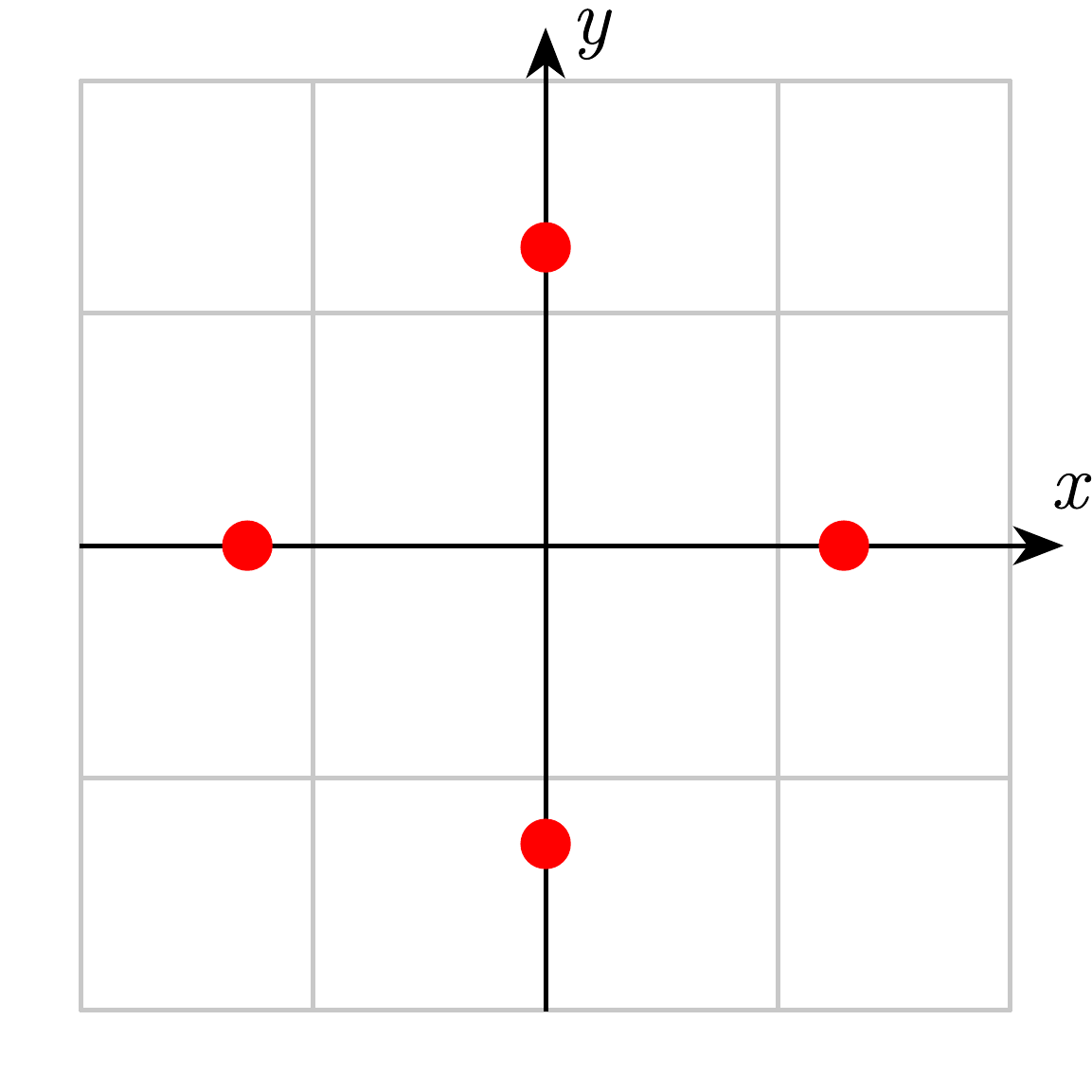}}\hspace{5pt}
    \subfloat[ReSpeaker Core]{\includegraphics[width=0.4\linewidth]{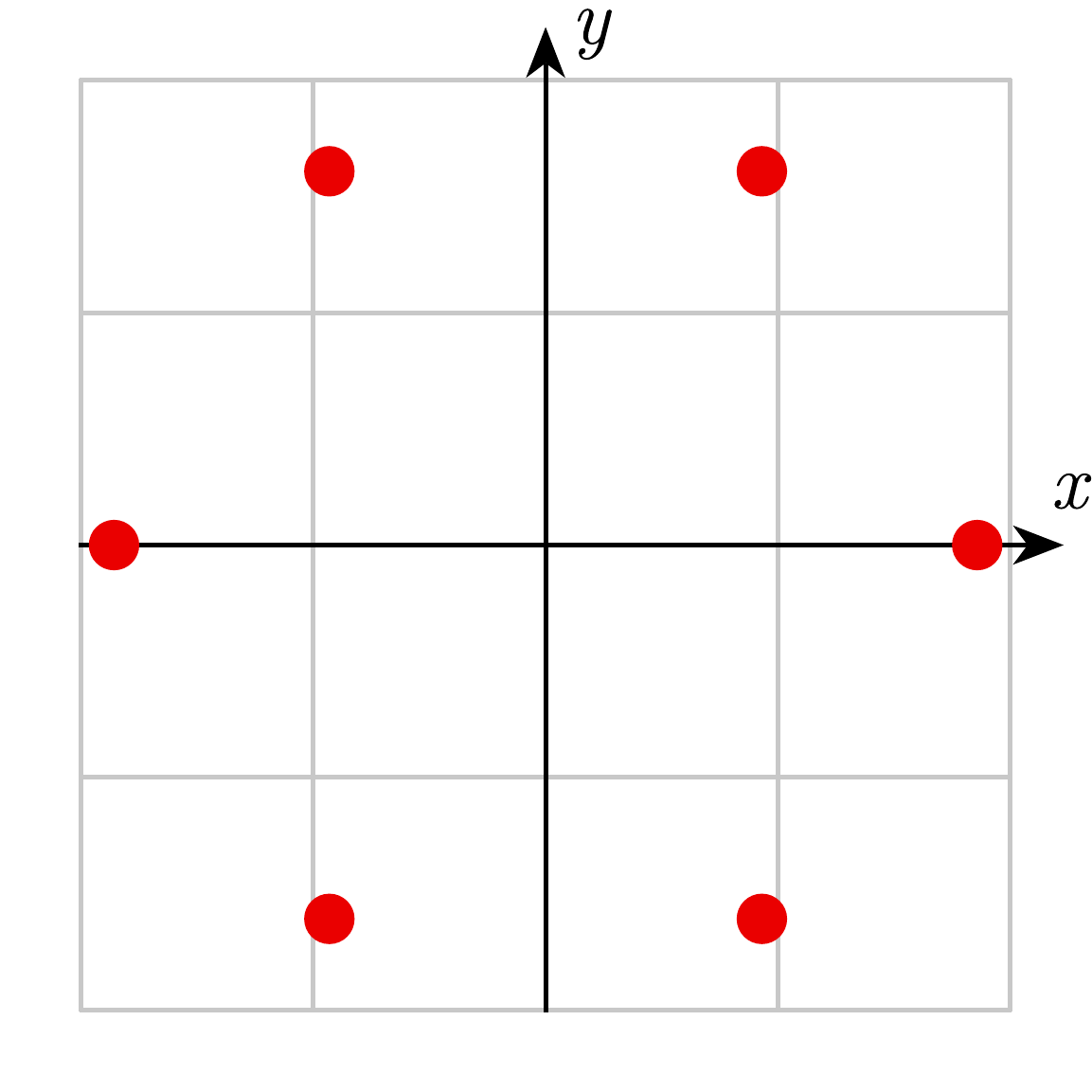}}\\
    \subfloat[MiniDSP UMA]{\includegraphics[width=0.4\linewidth]{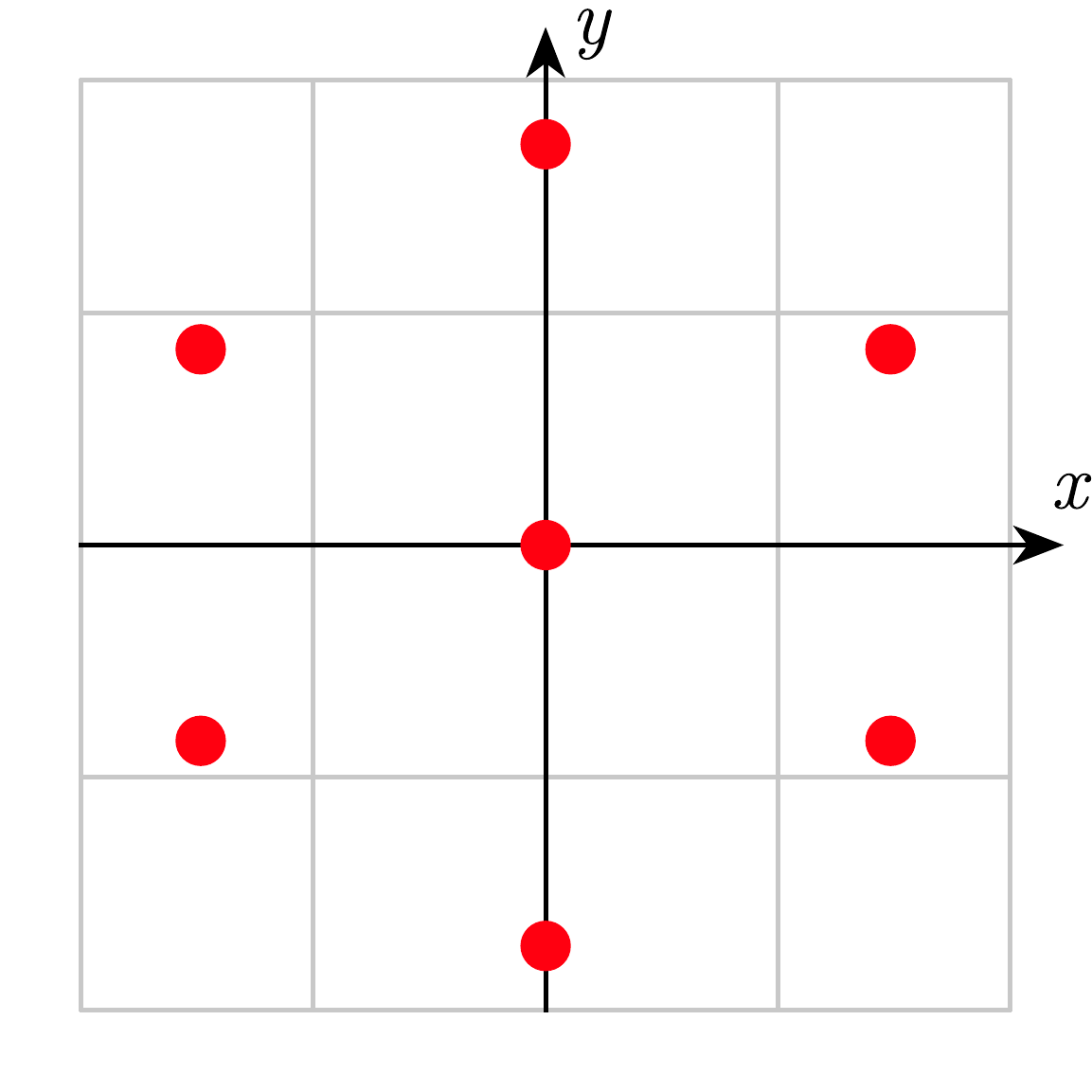}}\hspace{5pt}
    \subfloat[Matrix Creator]{\includegraphics[width=0.4\linewidth]{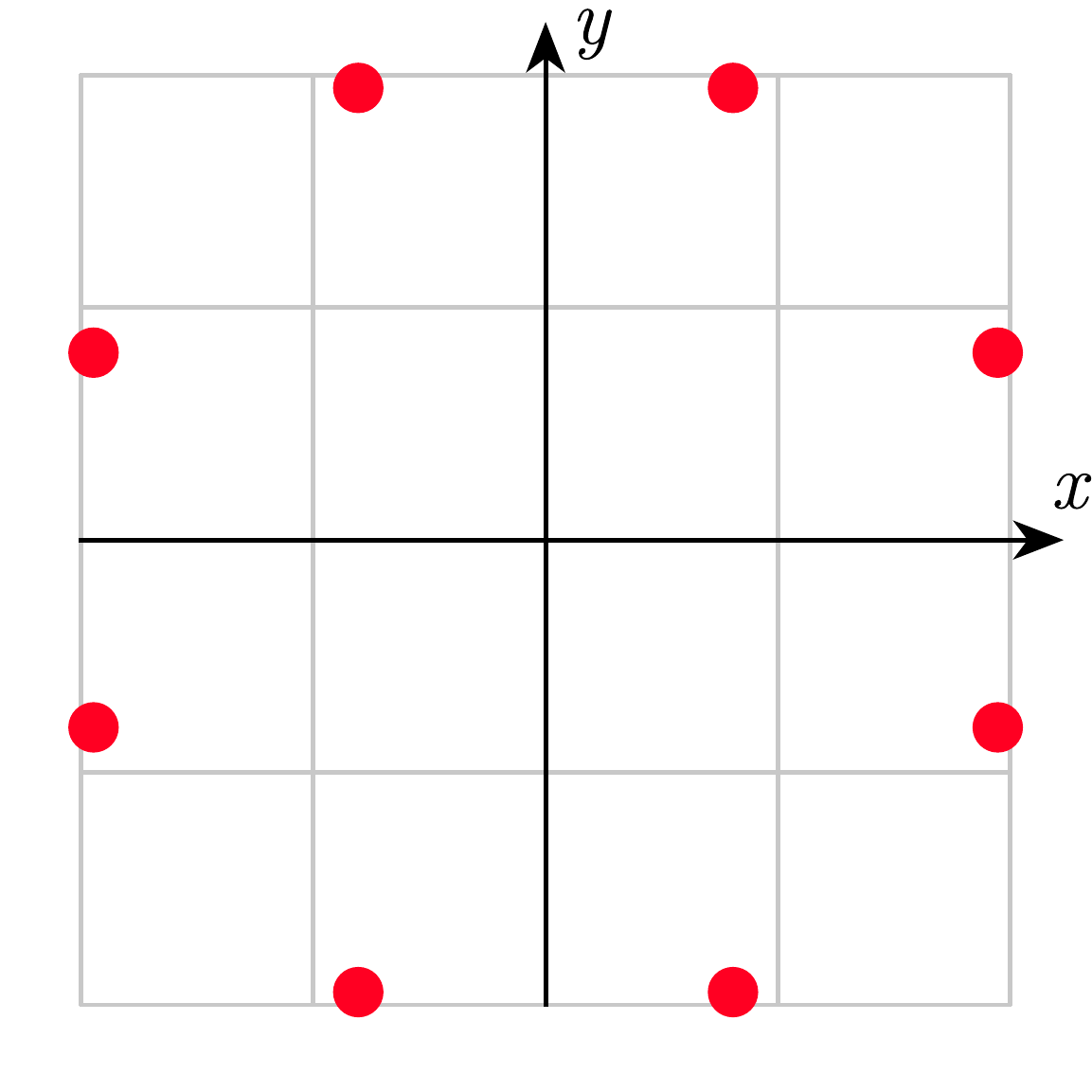}}
    \caption{Microphone array geometries}
    \label{fig:micarray}
\end{figure}

We validate the proposed method with multiple commercial microphone arrays configurations: 1) ReSpeaker USB ($M=4$) (USB) \cite{sudharsan2019ai}, 2) Respeaker Core ($M=6$) (Core), 3) MiniDSP UMA ($M=7$) (UMA) \cite{agarwal2018opportunistic}, and 4) Matrix Creator ($M=8$) (MC) \cite{haider2019system}.
Their respective geometry is shown in Figure \ref{fig:micarray}, and the exact positions of each microphone are listed in Table \ref{tab:positions}.

\begin{table}[!ht]
    \centering
    \renewcommand{\arraystretch}{1.2}
    \caption{Microphone positions for microphone arrays (in m)}
    \begin{tabular}{c|cccc}
        \hline
        \hline
        Coordinates & $m$ & $x$ & $y$ & $z$ \\
        \hline
        \multirow{4}{*}{ReSpeaker USB} & 1 & -0.0320 & +0.0000 & +0.0000 \\
        & 2 & +0.0000 & -0.0320 & +0.0000 \\
        & 3 & +0.0320 & +0.0000 & +0.0000 \\
        & 4 & +0.0000 & +0.0320 & +0.0000 \\
        \hline
        \multirow{6}{*}{ReSpeaker Core} & 1 & -0.0232 & +0.0401 & +0.0000 \\
        & 2 & -0.0463 & +0.0000 & +0.0000 \\
        & 3 & -0.0232 & -0.0401 & +0.0000 \\
        & 4 & +0.0232 & -0.0401 & +0.0000 \\        
        & 5 & +0.0463 & +0.0000 & +0.0000 \\
        & 6 & +0.0232 & +0.0401 & +0.0000 \\        
        \hline
        \multirow{7}{*}{MiniDSP UMA} & 1 & +0.0000 & +0.0000 & +0.0000 \\
        & 2 & +0.0000 & +0.0430 & +0.0000 \\
        & 3 & +0.0370 & +0.0210 & +0.0000 \\
        & 4 & +0.0370 & -0.0210 & +0.0000 \\ 
        & 5 & +0.0000 & -0.0430 & +0.0000 \\         
        & 6 & -0.0370 & -0.0210 & +0.0000 \\
        & 7 & -0.0370 & +0.0210 & +0.0000 \\
        \hline
        \multirow{8}{*}{Matrix Creator} & 1 & +0.0201 & -0.0485 & +0.0000 \\
        & 2 & -0.0201 & -0.0485 & +0.0000 \\
        & 3 & -0.0485 & -0.0201 & +0.0000 \\
        & 4 & -0.0485 & +0.0201 & +0.0000 \\ 
        & 5 & -0.0201 & +0.0485 & +0.0000 \\         
        & 6 & +0.0201 & +0.0485 & +0.0000 \\
        & 7 & +0.0485 & +0.0201 & +0.0000 \\          
        & 8 & +0.0485 & -0.0201 & +0.0000 \\                  
        \hline        
        \hline
    \end{tabular}
    \label{tab:positions}
\end{table}

To validate that the proposed SMP-PHAT is as accurate as the SRP-PHAT algorithm, we simulate multiple acoustic conditions using the image method.
The microphone array is positioned randomly at a height of 1m and the white noise sound source at a height of 2m in a 10m $\times$ 10m $\times$ 3m rectangular room.
The reverberation time (RT60) varies from $0.2$ sec to $0.5$ sec, the speed of sound corresponds to $343$ m/sec, and the sample rate is set to $f_S = 16000$ samples/sec.
For each microphone array, there are $L=1000$ simulations, indexed by $l \in \{1, 2, \dots, L\}$, where we compare the predicted DoA using SRP-PHAT and SMP-PHAT (denoted as $\mathbf{u}_l^*$) with the theoretical DoA (denoted by the unit vector $\bar{\mathbf{u}}_l$), and compute the mean average error (MAE) as follows:
\begin{equation}
    \mathrm{MAE} = \frac{1}{L}\sum_{l=1}^{L}{\arccos{\{\mathbf{u}_l^* \cdot \bar{\mathbf{u}}_l\}}}
\end{equation}

The number of computations (IFFTs, lookups and additions) are obtained using the equations introduced in Table \ref{tab:computations} once the parameter $Q$ is found using Algorithm \ref{alg:merging}.
The computing time is also obtained by running the SRP-PHAT and SMP-PHAT algorithms on Raspberry Pi 4 (RP4) and Raspberry Pi Zero (RPZ) devices.
The algorithms are written in the C language\footnote{github.com/francoisgrondin/smpphat} and compiled with the GNU Compiler Collection \cite{stallman1999using}, and the FFTW library \cite{johnson2006modified} is used to perform the IFFT operation.
These results are summarized in Table \ref{tab:results}.
Results demonstrate that the proposed SMP-PHAT is as accurate as SRP-PHAT, yet the number of computations is reduced by a factor between $-20\%$ and $-39\%$, depending on the array geometry.
This is directly observed while running on low-cost hardware, as the execution time goes down by a factor between $-28\%$ and $-39\%$, and $-23\%$ and $-34\%$, for the RP4 and RPZ, respectively.
This clearly confirms the improved efficiency of the proposed SMP-PHAT over the traditional SRP-PHAT method.

\begin{table}[]
    \centering
    \renewcommand{\arraystretch}{1.2}
    \caption{Mean Average Error (MAE, in degrees -- less is better), number of operations and execution time (in msec -- less is better) on different hardware, with SRP-PHAT (SRP) compared to SMP-PHAT (SMP)}
    \begin{tabular}{cc|cccc}
        \hline
        \hline
        \multicolumn{2}{c|}{Microphone Array} & USB & Core & UMA & MC \\
        \hline
        \multirow{5}{*}{Parameters} & $M$ & 4 & 6 & 7 & 8 \\
                                    & $P$ & 6 & 15 & 21 & 28 \\
                                    & $Q$ & 4 & 9 & 12 & 16 \\
                                    & $N$ & 512 & 512 & 512 & 512 \\
                                    & $I$ & 1321 & 1321 & 1321 & 321 \\
                                    & $k$ & 4 & 4 & 4 & 4 \\
        \hline
        \multirow{3}{*}{MAE}        & SRP       & 16.23 & 14.81 & 15.92 & 14.17 \\
                                    & SMP       & 16.23 & 14.81 & 15.92 & 14.17 \\
                                    & $\Delta$  & 0.00 & 0.00 & 0.00 & 0.00 \\
        \hline
        \multirow{3}{*}{IFFTs}      & SRP       & 6 & 15 & 21 & 28 \\
                                    & SMP       & 4 & 9 & 12 & 16 \\
                                    & $\Delta$  & -33\% & -40\% & -43\% & -43\% \\
        \hline
        \multirow{3}{*}{Lookups}    & SRP      & 7926 & 19815 & 27741 & 36988 \\
                                    & SMP      & 5284 & 11889 & 15852 & 21136 \\
                                    & $\Delta$  & -33\% & -40\% & -43\% & -43\% \\
        \hline
        \multirow{3}{*}{Additions}  & SRP       & 7926 & 19815 & 27741 & 36988 \\
                                    & SMP       & 6312 & 14973 & 20478 & 27304 \\
                                    & $\Delta$  & -20\% & -24\% & -26\% & -26\% \\
        \hline
        \multirow{3}{*}{Time (RP4)} & SRP       & 0.1410 & 0.3623 & 0.4690 & 0.6224 \\
                                    & SMP       & 0.1014 & 0.2214 & 0.2953 & 0.3912 \\
                                    & $\Delta$  & -28\% & -39\% & -37\% & -37\% \\
        \hline
        \multirow{3}{*}{Time (RPZ)} & SRP       & 2.9639 & 7.4741 & 10.6384 & 14.5013 \\
                                    & SMP       & 2.2966 & 5.1800 & 7.2243 & 9.5784 \\
                                    & $\Delta$  & -23\% & -31\% & -32\% & -34\% \\
        \hline
        \hline
    \end{tabular}
    \label{tab:results}
\end{table}

\section{Conclusion}
\label{sec:conclusion}

This paper introduces the SMP-PHAT method, which is an efficient method to perform DoA estimation for small microphone arrays with symmetrical geometries.
Results show that the proposed approach is as accurate as SRP-PHAT, yet reduces the computational load by a factor of down to $-39\%$ and $-34\%$ when it runs on low-cost hardware like Raspberry Pi 4 and Raspberry Pi Zero, respectively.
This makes SMP-PHAT well-suited for embedded processing on portable devices with limited computing ressources.
In future work, the method could be adapted for multiple sound source localization, and combine with sound source tracking algorithms.
Time-frequency masking could also be introduced to filter undesirable sound sources and focus on sources of interest, such as speech.

\clearpage

\bibliographystyle{IEEEtran}
\bibliography{mybib}

\begin{thebibliography}{10}
\providecommand{\url}[1]{#1}
\csname url@samestyle\endcsname
\providecommand{\newblock}{\relax}
\providecommand{\bibinfo}[2]{#2}
\providecommand{\BIBentrySTDinterwordspacing}{\spaceskip=0pt\relax}
\providecommand{\BIBentryALTinterwordstretchfactor}{4}
\providecommand{\BIBentryALTinterwordspacing}{\spaceskip=\fontdimen2\font plus
\BIBentryALTinterwordstretchfactor\fontdimen3\font minus
  \fontdimen4\font\relax}
\providecommand{\BIBforeignlanguage}[2]{{%
\expandafter\ifx\csname l@#1\endcsname\relax
\typeout{** WARNING: IEEEtran.bst: No hyphenation pattern has been}%
\typeout{** loaded for the language `#1'. Using the pattern for}%
\typeout{** the default language instead.}%
\else
\language=\csname l@#1\endcsname
\fi
#2}}
\providecommand{\BIBdecl}{\relax}
\BIBdecl

\bibitem{barker2015third}
J.~Barker, R.~Marxer, E.~Vincent, and S.~Watanabe, ``The third {CHiME} speech
  separation and recognition challenge: Dataset, task and baselines,'' in
  \emph{Proc. IEEE Workshop on ASRU}, 2015, pp. 504--511.

\bibitem{pertila2018multichannel}
P.~Pertil{\"a}, A.~Brutti, P.~Svaizer, and M.~Omologo, ``Multichannel source
  activity detection, localization, and tracking,'' \emph{Audio source
  separation and speech enhancement}, pp. 47--64, 2018.

\bibitem{wang2018multi}
Z.-Q. Wang, J.~Le~Roux, and J.~R. Hershey, ``Multi-channel deep clustering:
  Discriminative spectral and spatial embeddings for speaker-independent speech
  separation,'' in \emph{Proc. IEEE ICASSP}, 2018, pp. 1--5.

\bibitem{drude2019unsupervised}
L.~Drude, D.~Hasenklever, and R.~Haeb-Umbach, ``Unsupervised training of a deep
  clustering model for multichannel blind source separation,'' in \emph{Proc.
  IEEE ICASSP}, 2019, pp. 695--699.

\bibitem{grondin2020gev}
F.~Grondin, J.-S. Lauzon, J.~Vincent, and F.~Michaud, ``{GEV} beamforming
  supported by {DOA}-based masks generated on pairs of microphones,'' in
  \emph{Proc. Interspeech}, 2020, pp. 3341--3345.

\bibitem{wood2017blind}
S.~U. Wood, J.~Rouat, S.~Dupont, and G.~Pironkov, ``Blind speech separation and
  enhancement with {GCC-NMF},'' \emph{IEEE/ACM Trans. Audio, Speech, Language
  Process.}, vol.~25, no.~4, pp. 745--755, 2017.

\bibitem{pertila2015distant}
P.~Pertil{\"a} and J.~Nikunen, ``Distant speech separation using predicted
  time--frequency masks from spatial features,'' \emph{Speech commun.},
  vol.~68, pp. 97--106, 2015.

\bibitem{do2007real}
H.~Do, H.~F. Silverman, and Y.~Yu, ``A real-time {SRP-PHAT} source location
  implementation using stochastic region contraction ({SRC}) on a
  large-aperture microphone array,'' in \emph{Proc. IEEE ICASSP}, vol.~1, 2007,
  pp. 121--124.

\bibitem{valin2004localization}
J.-M. Valin, F.~Michaud, B.~Hadjou, and J.~Rouat, ``Localization of
  simultaneous moving sound sources for mobile robot using a frequency-domain
  steered beamformer approach,'' in \emph{Proc. IEEE ICRA}, vol.~1, 2004, pp.
  1033--1038.

\bibitem{valin2007robust}
J.-M. Valin, F.~Michaud, and J.~Rouat, ``Robust localization and tracking of
  simultaneous moving sound sources using beamforming and particle filtering,''
  \emph{Robot. Auton. Syst.}, vol.~55, no.~3, pp. 216--228, 2007.

\bibitem{brandstein1997robust}
M.~S. Brandstein and H.~F. Silverman, ``A robust method for speech signal
  time-delay estimation in reverberant rooms,'' in \emph{Proc. IEEE ICASSP},
  vol.~1, 1997, pp. 375--378.

\bibitem{cohen2001speech}
I.~Cohen and B.~Berdugo, ``Speech enhancement for non-stationary noise
  environments,'' \emph{Signal Process.}, vol.~81, no.~11, pp. 2403--2418,
  2001.

\bibitem{grondin2015time}
F.~Grondin and F.~Michaud, ``Time difference of arrival estimation based on
  binary frequency mask for sound source localization on mobile robots,'' in
  \emph{Proc. IEEE/RSJ IROS}, 2015, pp. 6149--6154.

\bibitem{grondin2016noise}
------, ``Noise mask for {TDOA} sound source localization of speech on mobile
  robots in noisy environments,'' in \emph{Proc. IEEE ICRA}, 2016, pp.
  4530--4535.

\bibitem{pertila2017robust}
P.~Pertil{\"a} and E.~Cakir, ``Robust direction estimation with convolutional
  neural networks based steered response power,'' in \emph{Proc. IEEE ICASSP},
  2017, pp. 6125--6129.

\bibitem{wang2018robust}
Z.-Q. Wang, X.~Zhang, and D.~Wang, ``Robust speaker localization guided by deep
  learning-based time-frequency masking,'' \emph{IEEE/ACM Trans. Audio, Speech,
  Language Process.}, vol.~27, no.~1, pp. 178--188, 2018.

\bibitem{grondin2013manyears}
F.~Grondin, D.~L{\'e}tourneau, F.~Ferland, V.~Rousseau, and F.~Michaud, ``The
  {ManyEars} open framework,'' \emph{Auton. Robots}, vol.~34, no.~3, pp.
  217--232, 2013.

\bibitem{grondin2019lightweight}
F.~Grondin and F.~Michaud, ``Lightweight and optimized sound source
  localization and tracking methods for open and closed microphone array
  configurations,'' \emph{Robot. Auton. Syst.}, vol. 113, pp. 63--80, 2019.

\bibitem{grondin2021odas}
F.~Grondin, D.~L{\'e}tourneau, C.~Godin, J.-S. Lauzon, J.~Vincent, S.~Michaud,
  S.~Faucher, and F.~Michaud, ``{ODAS}: Open embedded audition system,''
  \emph{arXiv preprint arXiv:2103.03954}, 2021.

\bibitem{han2005esprit}
F.-M. Han and X.-D. Zhang, ``An {ESPRIT}-like algorithm for coherent doa
  estimation,'' \emph{IEEE Antennas Wirel. Propag. Lett.}, vol.~4, pp.
  443--446, 2005.

\bibitem{schmidt1986multiple}
R.~Schmidt, ``Multiple emitter location and signal parameter estimation,''
  \emph{IEEE Trans. Antennas Propag.}, vol.~34, no.~3, pp. 276--280, 1986.

\bibitem{ishi2009evaluation}
C.~T. Ishi, O.~Chatot, H.~Ishiguro, and N.~Hagita, ``Evaluation of a
  {MUSIC}-based real-time sound localization of multiple sound sources in real
  noisy environments,'' in \emph{Proc. IEEE/RSJ IROS}, 2009, pp. 2027--2032.

\bibitem{nakamura2009intelligent}
K.~Nakamura, K.~Nakadai, F.~Asano, Y.~Hasegawa, and H.~Tsujino, ``Intelligent
  sound source localization for dynamic environments,'' in \emph{Proc. IEEE/RSJ
  IROS}, 2009, pp. 664--669.

\bibitem{nakamura2011intelligent}
K.~Nakamura, K.~Nakadai, F.~Asano, and G.~Ince, ``Intelligent sound source
  localization and its application to multimodal human tracking,'' in
  \emph{Proc. IEEE/RSJ IROS}, 2011, pp. 143--148.

\bibitem{nakadai2012robot}
K.~Nakadai, G.~Ince, K.~Nakamura, and H.~Nakajima, ``Robot audition for dynamic
  environments,'' in \emph{Proc. IEEE ICSPCC}, 2012, pp. 125--130.

\bibitem{nakamura2012real}
K.~Nakamura, K.~Nakadai, and G.~Ince, ``Real-time super-resolution sound source
  localization for robots,'' in \emph{Proc. IEEE/RSJ IROS}, 2012, pp. 694--699.

\bibitem{chakrabarty2017broadband}
S.~Chakrabarty and E.~A. Habets, ``Broadband {DOA} estimation using
  convolutional neural networks trained with noise signals,'' in \emph{Proc.
  IEEE WASPAA}, 2017, pp. 136--140.

\bibitem{chakrabarty2019multi}
------, ``Multi-speaker {DOA} estimation using deep convolutional networks
  trained with noise signals,'' \emph{IEEE J Sel Top Signal Process}, vol.~13,
  no.~1, pp. 8--21, 2019.

\bibitem{adavanne2018sound}
S.~Adavanne, A.~Politis, J.~Nikunen, and T.~Virtanen, ``Sound event
  localization and detection of overlapping sources using convolutional
  recurrent neural networks,'' \emph{IEEE J Sel Top Signal Process}, vol.~13,
  no.~1, pp. 34--48, 2018.

\bibitem{adavanne2018direction}
S.~Adavanne, A.~Politis, and T.~Virtanen, ``Direction of arrival estimation for
  multiple sound sources using convolutional recurrent neural network,'' in
  \emph{Proc. EUSIPCO}, 2018, pp. 1462--1466.

\bibitem{grondin2019sound}
F.~Grondin, J.~Glass, I.~Sobieraj, and M.~D. Plumbley, ``Sound event
  localization and detection using crnn on pairs of microphones,'' \emph{arXiv
  preprint arXiv:1910.10049}, 2019.

\bibitem{li2018online}
Q.~Li, X.~Zhang, and H.~Li, ``Online direction of arrival estimation based on
  deep learning,'' in \emph{Proc. IEEE ICASSP}.\hskip 1em plus 0.5em minus
  0.4em\relax IEEE, 2018, pp. 2616--2620.

\bibitem{mcmanus2021raspberry}
S.~McManus and M.~Cook, \emph{Raspberry Pi for dummies}.\hskip 1em plus 0.5em
  minus 0.4em\relax John Wiley \& Sons, 2021.

\bibitem{grimmett2016getting}
R.~Grimmett, \emph{Getting Started with Raspberry Pi Zero}.\hskip 1em plus
  0.5em minus 0.4em\relax Packt Publishing Ltd, 2016.

\bibitem{sudharsan2019ai}
B.~Sudharsan, S.~P. Kumar, and R.~Dhakshinamurthy, ``{AI} vision: Smart speaker
  design and implementation with object detection custom skill and advanced
  voice interaction capability,'' in \emph{Proc. IEEE ICoAC}, 2019, pp.
  97--102.

\bibitem{agarwal2018opportunistic}
A.~Agarwal, M.~Jain, P.~Kumar, and S.~Patel, ``Opportunistic sensing with mic
  arrays on smart speakers for distal interaction and exercise tracking,'' in
  \emph{Proc. IEEE ICASSP}, 2018, pp. 6403--6407.

\bibitem{haider2019system}
F.~Haider and S.~Luz, ``A system for real-time privacy preserving data
  collection for ambient assisted living.'' in \emph{Proc. Interspeech}, 2019,
  pp. 2374--2375.

\bibitem{stallman1999using}
R.~M. Stallman, \emph{Using and porting the GNU compiler collection}.\hskip 1em
  plus 0.5em minus 0.4em\relax Free Software Foundation, 1999, vol.~86.

\bibitem{johnson2006modified}
S.~G. Johnson and M.~Frigo, ``A modified split-radix {FFT} with fewer
  arithmetic operations,'' \emph{IEEE Trans. Signal Process.}, vol.~55, no.~1,
  pp. 111--119, 2006.

\end{thebibliography}

\end{document}